\documentclass[a4paper,11pt]{article}
\usepackage{pos}

\usepackage{mciteplus}
\usepackage{adjustbox}
\usepackage{subcaption}
\usepackage{tikz}
\usepackage{tikz-feynman}
\usepackage{slashed}
\usepackage{mathtools}
\usepackage[capitalize]{cleveref}

\newcommand{\rmd}{\mathrm{d}}
\newcommand{\ep}{\epsilon}
\newcommand{\progname}[1]{\textsc{#1}}
\DeclareMathOperator*{\MaxCut}{MaxCut}

\allowdisplaybreaks

\title{Four-loop two-mass tadpoles and the \texorpdfstring{$\rho$}{rho} parameter}

\author[a,b]{Samuel Abreu}
\author*[a]{Arnd Behring}
\author[a,b]{Andrew McLeod}
\author[a,c]{Ben Page}

\affiliation[a]{Theoretical Physics Department, CERN, \\
  1211 Geneva 23, Switzerland}
\affiliation[b]{Higgs Centre for Theoretical Physics,
  School of Physics and Astronomy, \\
  The University of Edinburgh,
  Edinburgh EH9 3FD, Scotland, UK}
\affiliation[c]{Department of Physics and Astronomy, Ghent University, \\
  9000 Ghent, Belgium}

\emailAdd{arnd.behring@cern.ch}

\abstract{We calculate four-loop QCD corrections to the electroweak $\rho$
  parameter with a non-vanishing $b$ quark mass. At three loops, it was
  observed that elliptic integrals contribute to this observable. This prompts
  the question of which classes of functions appear at the next order. We
  report on the status of our calculation with a focus on the mathematical
  structures that emerge at four loops.}

\FullConference{Loops and Legs in Quantum Field Theory (LL2024)\\
  14--19 April 2024\\
  Wittenberg, Germany}

\begin{document}
\definecolor{mt}{HTML}{CC0000}
\definecolor{mb}{HTML}{00AA00}
\definecolor{m0}{HTML}{000000}
\definecolor{m3}{HTML}{0000CC}
\definecolor{cut}{HTML}{0000FF}

\tikzset{B00/.pic={code={
  \begin{feynman}
    \vertex (i1) at (-1.5, 0.3);
    \vertex (i2) at (-1.5,-0.3);
    \vertex (o1) at ( 1.5, 0.3);
    \vertex (o2) at ( 1.5,-0.3);
    \vertex (v1) at (-1.0, 0.0);
    \vertex (v2) at ( 1.0, 0.0);
    \diagram*{
      (v1) --[plain,m0,out=  80,in= 100,looseness=1.0] (v2)
           --[plain,m0,out=-100,in=- 80,looseness=1.0] (v1),
      (i1) --[plain,m0] (v1) --[plain] (i2),
      (o1) --[plain,m0] (v2) --[plain] (o2)
    };
  \end{feynman}
}}}

\tikzset{Bb0/.pic={code={
  \begin{feynman}
    \vertex (i1) at (-1.5, 0.3);
    \vertex (i2) at (-1.5,-0.3);
    \vertex (o1) at ( 1.5, 0.3);
    \vertex (o2) at ( 1.5,-0.3);
    \vertex (v1) at (-1.0, 0.0);
    \vertex (v2) at ( 1.0, 0.0);
    \diagram*{
      (v1) --[plain,mb,out=  80,in= 100,looseness=1.0] (v2)
           --[plain,m0,out=-100,in=- 80,looseness=1.0] (v1),
      (i1) --[plain,m0] (v1) --[plain] (i2),
      (o1) --[plain,m0] (v2) --[plain] (o2)
    };
  \end{feynman}
}}}

\tikzset{Bt0/.pic={code={
  \begin{feynman}
    \vertex (i1) at (-1.5, 0.3);
    \vertex (i2) at (-1.5,-0.3);
    \vertex (o1) at ( 1.5, 0.3);
    \vertex (o2) at ( 1.5,-0.3);
    \vertex (v1) at (-1.0, 0.0);
    \vertex (v2) at ( 1.0, 0.0);
    \diagram*{
      (v1) --[plain,mt,out=  80,in= 100,looseness=1.0] (v2)
           --[plain,m0,out=-100,in=- 80,looseness=1.0] (v1),
      (i1) --[plain,m0] (v1) --[plain] (i2),
      (o1) --[plain,m0] (v2) --[plain] (o2)
    };
  \end{feynman}
}}}

\tikzset{Bbb/.pic={code={
  \begin{feynman}
    \vertex (i1) at (-1.5, 0.3);
    \vertex (i2) at (-1.5,-0.3);
    \vertex (o1) at ( 1.5, 0.3);
    \vertex (o2) at ( 1.5,-0.3);
    \vertex (v1) at (-1.0, 0.0);
    \vertex (v2) at ( 1.0, 0.0);
    \diagram*{
      (v1) --[plain,mb,out=  80,in= 100,looseness=1.0] (v2)
           --[plain,mb,out=-100,in=- 80,looseness=1.0] (v1),
      (i1) --[plain,m0] (v1) --[plain] (i2),
      (o1) --[plain,m0] (v2) --[plain] (o2)
    };
  \end{feynman}
}}}

\tikzset{Btt/.pic={code={
  \begin{feynman}
    \vertex (i1) at (-1.5, 0.3);
    \vertex (i2) at (-1.5,-0.3);
    \vertex (o1) at ( 1.5, 0.3);
    \vertex (o2) at ( 1.5,-0.3);
    \vertex (v1) at (-1.0, 0.0);
    \vertex (v2) at ( 1.0, 0.0);
    \diagram*{
      (v1) --[plain,mt,out=  80,in= 100,looseness=1.0] (v2)
           --[plain,mt,out=-100,in=- 80,looseness=1.0] (v1),
      (i1) --[plain,m0] (v1) --[plain] (i2),
      (o1) --[plain,m0] (v2) --[plain] (o2)
    };
  \end{feynman}
}}}

\tikzset{Bbt/.pic={code={
  \begin{feynman}
    \vertex (i1) at (-1.5, 0.3);
    \vertex (i2) at (-1.5,-0.3);
    \vertex (o1) at ( 1.5, 0.3);
    \vertex (o2) at ( 1.5,-0.3);
    \vertex (v1) at (-1.0, 0.0);
    \vertex (v2) at ( 1.0, 0.0);
    \diagram*{
      (v1) --[plain,mb,out=  80,in= 100,looseness=1.0] (v2)
           --[plain,mt,out=-100,in=- 80,looseness=1.0] (v1),
      (i1) --[plain,m0] (v1) --[plain] (i2),
      (o1) --[plain,m0] (v2) --[plain] (o2)
    };
  \end{feynman}
}}}

\tikzset{Sbbb/.pic={code={
  \begin{feynman}
    \vertex (i1) at (-1.5, 0.3);
    \vertex (i2) at (-1.5,-0.3);
    \vertex (o1) at ( 1.5, 0.3);
    \vertex (o2) at ( 1.5,-0.3);
    \vertex (v1) at (-1.0, 0.0);
    \vertex (v2) at ( 1.0, 0.0);
    \diagram*{
      (v1) --[plain,mb,out=  80,in= 100,looseness=1.0] (v2)
           --[plain,mb,out=-100,in=- 80,looseness=1.0] (v1)
           --[plain,mb] (v2),
      (i1) --[plain,m0] (v1) --[plain] (i2),
      (o1) --[plain,m0] (v2) --[plain] (o2)
    };
  \end{feynman}
}}}

\tikzset{SbbbOSt/.pic={code={
  \begin{feynman}
    \vertex (i1) at (-1.5, 0.0);
    \vertex (o1) at ( 1.5, 0.0);
    \vertex (v1) at (-1.0, 0.0);
    \vertex (v2) at ( 1.0, 0.0);
    \diagram*{
      (v1) --[plain,mb,out=  80,in= 100,looseness=1.0] (v2)
           --[plain,mb,out=-100,in=- 80,looseness=1.0] (v1)
           --[plain,mb] (v2),
      (i1) --[plain,mt] (v1),
      (o1) --[plain,mt] (v2)
    };
  \end{feynman}
}}}

\tikzset{banana3l/.pic={code={
  \begin{feynman}
    \vertex (v1) at (0,1);
    \vertex (v2) at (0,-1);
    \diagram*{
      (v1) --[plain,mb,out=0,in=0,looseness=1.7] (v2),
      (v1) --[plain,mb,out=0,in=0,looseness=0.8] (v2),
      (v1) --[plain,mb,out=180,in=180,looseness=0.8] (v2),
      (v1) --[plain,mt,out=180,in=180,looseness=1.7] (v2)
    };
  \end{feynman}
}}}

\tikzset{banana3ltwin/.pic={code={
  \begin{feynman}
    \vertex (v1) at (0,1);
    \vertex (v2) at (0,-1);
    \diagram*{
      (v1) --[plain,mt,out=0,in=0,looseness=1.7] (v2),
      (v1) --[plain,mt,out=0,in=0,looseness=0.8] (v2),
      (v1) --[plain,mt,out=180,in=180,looseness=0.8] (v2),
      (v1) --[plain,mb,out=180,in=180,looseness=1.7] (v2)
    };
  \end{feynman}
}}}

\tikzset{peint1/.pic={code={
  \begin{scope}[yshift=0.9cm]
  \begin{feynman}
    \vertex (v1) at (0,0);
    \vertex (v2) at (1,-1.73205);
    \vertex (v3) at (-1,-1.73025);
    \diagram*{
      (v1) --[plain,mb] (v2),
      (v1) --[plain,mb,out=-30,in=90] (v2),
      (v1) --[plain,mb,out=-90,in=150] (v2),
      (v2) --[plain,mt,out=-150,in=-30] (v3),
      (v3) --[plain,mb,out=90,in=-150] (v1),
      (v3) --[plain,m0,out=30,in=-90] (v1)
    };
  \end{feynman}
  \end{scope}
}}}

\tikzset{peint1twin/.pic={code={
  \begin{scope}[yshift=0.9cm]
  \begin{feynman}
    \vertex (v1) at (0,0);
    \vertex (v2) at (1,-1.73205);
    \vertex (v3) at (-1,-1.73025);
    \diagram*{
      (v1) --[plain,mt] (v2),
      (v1) --[plain,mt,out=-30,in=90] (v2),
      (v1) --[plain,mt,out=-90,in=150] (v2),
      (v2) --[plain,mb,out=-150,in=-30] (v3),
      (v3) --[plain,mt,out=90,in=-150] (v1),
      (v3) --[plain,m0,out=30,in=-90] (v1)
    };
  \end{feynman}
  \end{scope}
}}}

\tikzset{peint2/.pic={code={
  \begin{scope}[yshift=0.4cm]
  \begin{feynman}
    \vertex (v1) at (0,0);
    \vertex (v2) at (0,-1.5);
    \vertex (v3) at (0,0.7);
    \diagram*{
      (v1) --[plain,mt,out=180,in=180,looseness=1.7] (v2),
      (v1) --[plain,mb,out=180,in=180,looseness=0.8] (v2),
      (v1) --[plain,mb,out=0,in=0,looseness=0.8] (v2),
      (v1) --[plain,mb,out=0,in=0,looseness=1.7] (v2),
      (v1) --[plain,mb,out=0,in=0,looseness=1.7] (v3)
           --[plain,mb,out=180,in=180,looseness=1.7] (v1)
    };
  \end{feynman}
  \end{scope}
}}}

\tikzset{peint2twin/.pic={code={
  \begin{scope}[yshift=0.4cm]
  \begin{feynman}
    \vertex (v1) at (0,0);
    \vertex (v2) at (0,-1.5);
    \vertex (v3) at (0,0.7);
    \diagram*{
      (v1) --[plain,mb,out=180,in=180,looseness=1.7] (v2),
      (v1) --[plain,mt,out=180,in=180,looseness=0.8] (v2),
      (v1) --[plain,mt,out=0,in=0,looseness=0.8] (v2),
      (v1) --[plain,mt,out=0,in=0,looseness=1.7] (v2),
      (v1) --[plain,mt,out=0,in=0,looseness=1.7] (v3)
           --[plain,mt,out=180,in=180,looseness=1.7] (v1)
    };
  \end{feynman}
  \end{scope}
}}}

\tikzset{peint3/.pic={code={
  \begin{scope}[yshift=0.4cm]
  \begin{feynman}
    \vertex (v1) at (0,0);
    \vertex (v2) at (0,-1.5);
    \vertex (v3) at (0,0.7);
    \diagram*{
      (v1) --[plain,mt,out=180,in=180,looseness=1.7] (v2),
      (v1) --[plain,mb,out=180,in=180,looseness=0.8] (v2),
      (v1) --[plain,mb,out=0,in=0,looseness=0.8] (v2),
      (v1) --[plain,mb,out=0,in=0,looseness=1.7] (v2),
      (v1) --[plain,mt,out=0,in=0,looseness=1.7] (v3)
           --[plain,mt,out=180,in=180,looseness=1.7] (v1)
    };
  \end{feynman}
  \end{scope}
}}}

\tikzset{peint3twin/.pic={code={
  \begin{scope}[yshift=0.4cm]
  \begin{feynman}
    \vertex (v1) at (0,0);
    \vertex (v2) at (0,-1.5);
    \vertex (v3) at (0,0.7);
    \diagram*{
      (v1) --[plain,mb,out=180,in=180,looseness=1.7] (v2),
      (v1) --[plain,mt,out=180,in=180,looseness=0.8] (v2),
      (v1) --[plain,mt,out=0,in=0,looseness=0.8] (v2),
      (v1) --[plain,mt,out=0,in=0,looseness=1.7] (v2),
      (v1) --[plain,mb,out=0,in=0,looseness=1.7] (v3)
           --[plain,mb,out=180,in=180,looseness=1.7] (v1)
    };
  \end{feynman}
  \end{scope}
}}}

\tikzset{peint4/.pic={code={
  \begin{scope}[yshift=0.9cm]
  \begin{feynman}
    \vertex (v1) at (0,0);
    \vertex (v2) at (1,-1.73205);
    \vertex (v3) at (-1,-1.73025);
    \diagram*{
      (v1) --[plain,mt,out=-30,in=90] (v2),
      (v1) --[plain,mb,out=-90,in=150] (v2),
      (v2) --[plain,mb,out=-150,in=-30] (v3),
      (v2) --[plain,mb,out=150,in=30] (v3),
      (v3) --[plain,m0,out=90,in=-150] (v1),
      (v3) --[plain,m0,out=30,in=-90] (v1)
    };
  \end{feynman}
  \end{scope}
}}}

\tikzset{peint5/.pic={code={
  \begin{scope}[yshift=0.9cm]
  \begin{feynman}
    \vertex (v1) at (0,0);
    \vertex (v2) at (1,-1.73205);
    \vertex (v3) at (-1,-1.73025);
    \diagram*{
      (v1) --[plain,mt,out=-30,in=90] (v2),
      (v1) --[plain,mb,out=-90,in=150] (v2),
      (v2) --[plain,mb,out=-150,in=-30] (v3),
      (v2) --[plain,mb,out=150,in=30] (v3),
      (v3) --[plain,mt,out=90,in=-150] (v1),
      (v3) --[plain,mb,out=30,in=-90] (v1)
    };
  \end{feynman}
  \end{scope}
}}}

\tikzset{peint6/.pic={code={
  \begin{scope}[yshift=0.9cm]
  \begin{feynman}
    \vertex (v1) at (0,0);
    \vertex (v2) at (1,-1.73205);
    \vertex (v3) at (-1,-1.73025);
    \diagram*{
      (v1) --[plain,mb] (v2),
      (v1) --[plain,mt,out=-30,in=90] (v2),
      (v1) --[plain,mb,out=-90,in=150] (v2),
      (v2) --[plain,mb,out=-150,in=-30] (v3),
      (v3) --[plain,mt,out=90,in=-150] (v1),
      (v3) --[plain,m0,out=30,in=-90] (v1)
    };
  \end{feynman}
  \end{scope}
}}}

\tikzset{peint6twin/.pic={code={
  \begin{scope}[yshift=0.9cm]
  \begin{feynman}
    \vertex (v1) at (0,0);
    \vertex (v2) at (1,-1.73205);
    \vertex (v3) at (-1,-1.73025);
    \diagram*{
      (v1) --[plain,mt] (v2),
      (v1) --[plain,mb,out=-30,in=90] (v2),
      (v1) --[plain,mt,out=-90,in=150] (v2),
      (v2) --[plain,mt,out=-150,in=-30] (v3),
      (v3) --[plain,mb,out=90,in=-150] (v1),
      (v3) --[plain,m0,out=30,in=-90] (v1)
    };
  \end{feynman}
  \end{scope}
}}}

\tikzset{peint7/.pic={code={
  \begin{scope}[yshift=0.9cm]
  \begin{feynman}
    \vertex (v1) at (0,0);
    \vertex (v2) at (0,-2);
    \diagram*{
      (v1) --[plain,mt,out=180,in=180,looseness=1.7] (v2),
      (v1) --[plain,mb,out=180,in=180,looseness=0.8] (v2),
      (v1) --[plain,mb,out=0,in=0,looseness=0.8] (v2),
      (v1) --[plain,mb,out=0,in=0,looseness=1.7] (v2),
      (v1) --[plain,m0] (v2)
    };
  \end{feynman}
  \end{scope}
}}}

\tikzset{peint10/.pic={code={
  \begin{scope}[yshift=0.9cm]
  \begin{feynman}
    \vertex (v1) at (0,0);
    \vertex (v2) at (1,-1.73205);
    \vertex (v3) at (-1,-1.73025);
    \diagram*{
      (v1) --[plain,mb,out=-30,in=90] (v2),
      (v1) --[plain,mb,out=-90,in=150] (v2),
      (v2) --[plain,mt,out=-150,in=-30] (v3),
      (v2) --[plain,mb,out=150,in=30] (v3),
      (v3) --[plain,mt,out=90,in=-150] (v1),
      (v3) --[plain,mt,out=30,in=-90] (v1)
    };
  \end{feynman}
  \end{scope}
}}}

\tikzset{genus2/.pic={code={
  \begin{scope}[yshift=0.9cm]
  \begin{feynman}
    \vertex (v1) at (0,0);
    \vertex (v2) at (1,-1.73205);
    \vertex (v3) at (-1,-1.73025);
    \diagram*{
      (v1) --[plain,mb,out=-30,in=90] (v2),
      (v1) --[plain,mb,out=-90,in=150] (v2),
      (v2) --[plain,mt,out=-150,in=-30] (v3),
      (v2) --[plain,mb,out=150,in=30] (v3),
      (v3) --[plain,mt,out=90,in=-150] (v1),
      (v3) --[plain,m3,out=30,in=-90] (v1)
    };
  \end{feynman}
  \end{scope}
}}}

\maketitle

\section{Introduction}
The $\rho$~parameter is an important electroweak precision observable. It was
introduced by Ross and Veltman in 1975 \cite{Ross:1975fq} to distinguish the
Weinberg model, i.e., what turned into the modern Standard Model, from
alternative scenarios with extended Higgs sectors. Over the years, the
$\rho$~parameter has proven to be a valuable tool for precision tests of the
electroweak gauge sector of the Standard Model.
The $\rho$ parameter is defined as
\begin{align}
  \rho = \frac{m_W^2}{m_Z^2 \cos^2\theta_W}
  \,,
\end{align}
where $m_W$ and $m_Z$ are the masses of the $W$ and $Z$ boson, respectively,
and $\theta_W$ is the weak mixing angle.
At tree level in the Standard Model we have $\rho = 1$, but in New Physics
scenarios $\rho$ can differ from $1$ substantially. Also within the Standard
Model, higher-order contributions cause the $\rho$ parameter to deviate slightly
from its tree-level value.
There exists a long history of calculations that address these higher-order
corrections in the Standard Model, starting with the calculation of one-loop
corrections by Veltman \cite{Veltman:1977kh}. From there, one can compute
corrections at higher orders in the strong coupling $\alpha_s$
\cite{Djouadi:1987gn,Djouadi:1987di,Kniehl:1988ie,Kniehl:1989yc,%
Anselm:1993uq,Avdeev:1994db,Chetyrkin:1995ix,Grigo:2012ji,Blumlein:2018aeq,%
Abreu:2019fgk,Schroder:2005db,Chetyrkin:2006bj,Boughezal:2006xk} or the
electroweak coupling $\alpha$ \cite{vanderBij:1986hy,Barbieri:1992nz,%
Fleischer:1993ub,vanderBij:2000cg,Faisst:2003px}, or both \cite{Faisst:2003px}.

While calculating the three-loop QCD corrections with two non-vanishing quark
masses, it was observed that elliptic integrals appear \cite{Grigo:2012ji,%
Blumlein:2018aeq,Abreu:2019fgk}.
As also integrals over more complicated geometries such as hyperelliptic curves
and Calabi-Yau manifolds are known to appear in perturbative quantum field
theory \cite{Bourjaily:2022bwx}, it thus becomes natural to ask: what types of
integrals appear at four loops?

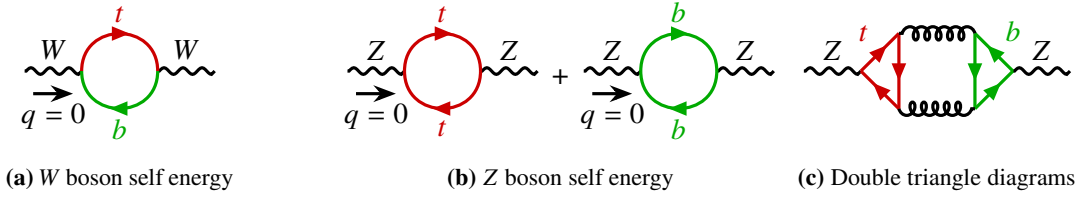
\begin{figure}
  \centering
  \subcaptionbox{$W$ boson self energy\label{fig:diag-W-1l}}[3cm]{%
    \begin{tikzpicture}[very thick]
      \useasboundingbox (-1.35,-1) rectangle (1.35,1);
      \begin{feynman}
        \vertex (i1) at (-1.25,0);
        \vertex (o1) at (1.25,0);
        \vertex (v1) at (-0.5,0);
        \vertex (v2) at (0.5,0);
        \diagram*{
          (i1) --[boson,momentum'={$q=0$},edge label={$W$}] (v1),
          (v2) --[boson,edge label={$W$}] (o1),
          (v1) --[fermion,mt,half left,looseness=1.7,
                  edge label={$\textcolor{mt}{t}$}] (v2),
          (v2) --[fermion,mb,half left,looseness=1.7,
                  edge label={$\textcolor{mb}{b}$}] (v1)
        };
      \end{feynman}
    \end{tikzpicture}
  }
  \hspace{3em}
  \subcaptionbox{$Z$ boson self energy\label{fig:diag-Z-1l}}{%
    \begin{adjustbox}{valign=m}
      \begin{tikzpicture}[very thick]
        \useasboundingbox (-1.35,-1) rectangle (1.35,1);
        \begin{feynman}
          \vertex (i1) at (-1.25,0);
          \vertex (o1) at (1.25,0);
          \vertex (v1) at (-0.5,0);
          \vertex (v2) at (0.5,0);
          \diagram*{
            (i1) --[boson,momentum'={$q=0$},edge label={$Z$}] (v1),
            (v2) --[boson,edge label={$Z$}] (o1),
            (v1) --[fermion,mt,half left,looseness=1.7,
                    edge label={$\textcolor{mt}{t}$}] (v2),
            (v2) --[fermion,mt,half left,looseness=1.7,
                    edge label={$\textcolor{mt}{t}$}] (v1)
          };
        \end{feynman}
      \end{tikzpicture}
    \end{adjustbox}
    $+$
    \begin{adjustbox}{valign=m}
      \begin{tikzpicture}[very thick]
        \useasboundingbox (-1.35,-1) rectangle (1.35,1);
        \begin{feynman}
          \vertex (i1) at (-1.25,0);
          \vertex (o1) at (1.25,0);
          \vertex (v1) at (-0.5,0);
          \vertex (v2) at (0.5,0);
          \diagram*{
            (i1) --[boson,momentum'={$q=0$},edge label={$Z$}] (v1),
            (v2) --[boson,edge label={$Z$}] (o1),
            (v1) --[fermion,mb,half left,looseness=1.7,
                    edge label={$\textcolor{mb}{b}$}] (v2),
            (v2) --[fermion,mb,half left,looseness=1.7,
                    edge label={$\textcolor{mb}{b}$}] (v1)
          };
        \end{feynman}
      \end{tikzpicture}
    \end{adjustbox}
  }
  \subcaptionbox{Double triangle diagrams\label{fig:double-triangle}}{%
    \begin{adjustbox}{valign=m}
      \begin{tikzpicture}[very thick]
        \useasboundingbox (-1.85,-1) rectangle (1.85,1);
        \begin{feynman}
          \vertex (i1) at (-1.75,0);
          \vertex (o1) at ( 1.75,0);
          \vertex (v1) at (-1.00,0);
          \vertex (v2) at (-0.50,0.5);
          \vertex (v3) at (-0.50,-0.5);
          \vertex (v4) at ( 1.00,0);
          \vertex (v5) at ( 0.50,0.5);
          \vertex (v6) at ( 0.50,-0.5);
          \diagram*{
            (i1) --[boson,edge label={$Z$}] (v1),
            (v4) --[boson,edge label={$Z$}] (o1),
            (v1) --[fermion,mt,edge label={\textcolor{mt}{$t$}}] (v2)
                 --[fermion,mt] (v3)
                 --[fermion,mt] (v1),
            (v4) --[fermion,mb,edge label'={\textcolor{mb}{$b$}}] (v5)
                 --[fermion,mb] (v6)
                 --[fermion,mb] (v4),
            (v5) --[gluon] (v2),
            (v3) --[gluon] (v6)
          };
        \end{feynman}
      \end{tikzpicture}
    \end{adjustbox}
  }
  \caption{Fermionic diagrams that contribute to the $W$ and $Z$ boson self
    energies.}
  \label{fig:one-loop-diags}
\end{figure}
Corrections to the $\rho$ parameter can be expressed as the difference between
the $Z$ and $W$ boson self energies,
\begin{align}
  \rho &= 1 + \Delta \rho
  \,, &
  \Delta \rho &= \frac{\Sigma_{Z}(q^2=0)}{m_Z^2} - \frac{\Sigma_W(q^2=0)}{m_W^2}
  \,.
\end{align}
In the Standard Model, the fermionic one-loop correction to the $W$ boson self
energy just consists of one Feynman diagram: a one-loop fermionic bubble with
up- and down-type quarks connecting the $W$ boson vertices.
The $Z$ boson self energy receives contributions from two diagrams, in which
only one flavour appears in the fermion loop. These diagrams are shown in
\cref{fig:diag-W-1l,fig:diag-Z-1l}. Due to the $\mathrm{SU}(2)_L$ structure of
the $W$ boson coupling to fermions, QCD corrections to the $W$ boson self
energy just add quark loops and gluons to the one-loop diagram. Similarly, the
majority of QCD corrections to $Z$ boson self energy are found by dressing the
one-loop diagrams with quark loops and gluons, except that, starting from
three-loop order, also double triangle diagrams appear (see
\cref{fig:double-triangle}).

The self energies in $\Delta \rho$ are evaluated for vanishing external
momentum, $q=0$, which means that the Feynman integrals are tadpole integrals.
They can be mapped to a small number of topologies (one topology each at one-,
two- and three-loop order, and two topologies at four-loop order). However,
this requires us to distinguish different mass labellings of each topology.
After generating the diagrams using \progname{QGRAF}~\cite{Nogueira:1991ex}, we
use \progname{FORM}~\cite{Ruijl:2017dtg} to insert Feynman rules, apply
projectors to extract the transversal part of the self energies and to simplify
the colour and Dirac algebra. We use integration-by-parts (IBP) relations
\cite{Chetyrkin:1980pr,Chetyrkin:1981qh,Tkachov:1981wb} via
\progname{Reduze2}~\cite{vonManteuffel:2012np} to reduce the integrals to a
smaller set of master integrals. The master integrals can be calculated using
the method of differential equations in the mass ratio $x = \frac{m_b}{m_t}$.
In order to obtain numerical values for the master integrals, we also use
\progname{AMFlow}~\cite{Liu:2017jxz,Liu:2021wks,Liu:2022mfb,Liu:2022chg}.

The numerical impact of the four-loop two-mass QCD corrections is expected to
be very small. Indeed, our main motivation for performing this calculation is
the question of which classes of functions are necessary to express the
solutions. Up to two-loop order, only harmonic polylogarithms are necessary to
express the two-mass corrections. At three-loop order, it was observed in
Ref.~\cite{Grigo:2012ji} that elliptic integrals appear. The explicit analytic
result in terms of iterated integrals was later obtained in
Refs.~\cite{Blumlein:2018aeq,Abreu:2019fgk}. It is thus natural to ask what
happens at four-loop order: will new classes of functions appear? Do we, for
example, encounter multiple elliptic curves, higher-genus curves, or
higher-dimensional geometries? We will address these questions in these
proceedings.

\section{Analysis}
After IBP reduction, we end up with 283 master integrals that can be organised
into 197 sectors. Many of these integrals are polylogarithmic, or can be
expressed in terms of polylogarithmic functions after rationalising a square
root. In order to identify which functions appear beyond multiple
polylogarithms, we need a criterion to narrow down the number of sectors that
we have to analyse in more detail. To this end, we analyse the differential
equations. In particular, we use the fact that new structures arise from the
homogeneous part of the differential equation.\footnote{Integrals can depend
on elliptic curves and more general geometries also through subtopologies.
Since we consider all sectors of master integrals that appear in the
calculation and since at this point we are only interested in surveying the new
mathematical structures that appear, we can restrict ourselves to investigating
only the maximal cuts.} That means that it is sufficient to consider the
differential equations in the limit $\ep \to 0$ and at the level of maximal
cuts. We derive the Picard-Fuchs operator by uncoupling the first-order
differential equation systems into a single higher-order differential equation
for one of the integrals of the sector. There are algorithms and packages that
automate this step \cite{Gerhold:2002,Zuercher:1994,AbramovZima:1996}. We then
try to find a factorisation of the differential operator, for example using
\progname{Maple}'s \texttt{DFactor} command \cite{vanHoeij:1997}. If it is
possible to find a factorisation into first-order factors, we can exclude that
new structures arise from this sector. However, if factors of order two or
higher appear, we take this as an indication that non-polylogarithmic
structures can emerge and that we need to analyse this sector in more detail.
We note that finding higher-order factors is not a sufficient criterion for
the appearance of elliptic integrals or generalisations thereof, as we will see
below.

\begin{figure}
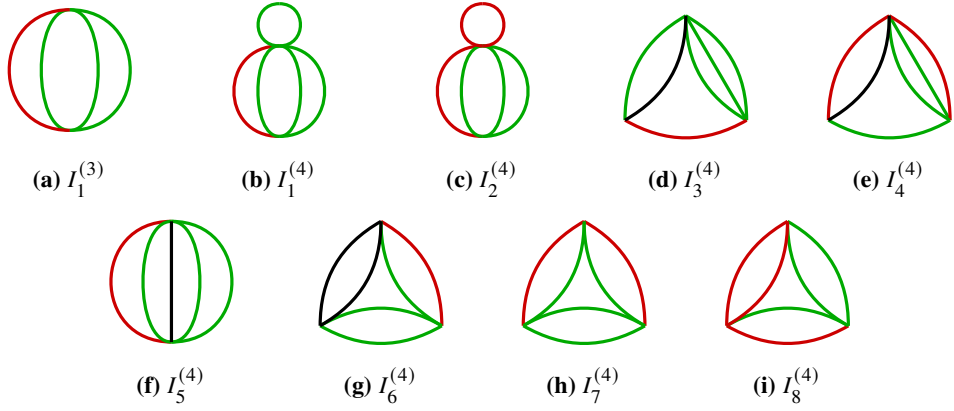

  \newcommand{\showdiag}[1]{%
    \tikz[very thick]{
      \draw pic[scale=0.8] at (0,0) {#1};
      \useasboundingbox (-1,-1) rectangle (1,1);
    }
  }
  \centering
  \subcaptionbox{$I_1^{(3)}$\label{fig:I3l1}}{\showdiag{banana3l}}
  \hspace{1em}
  \subcaptionbox{$I_1^{(4)}$\label{fig:I4l1}}{\showdiag{peint2}}
  \hspace{1em}
  \subcaptionbox{$I_2^{(4)}$\label{fig:I4l2}}{\showdiag{peint3}}
  \hspace{1em}
  \subcaptionbox{$I_3^{(4)}$\label{fig:I4l3}}{\showdiag{peint1}}
  \hspace{1em}
  \subcaptionbox{$I_4^{(4)}$\label{fig:I4l4}}{\showdiag{peint6}}
  \\
  \subcaptionbox{$I_5^{(4)}$\label{fig:I4l5}}{\showdiag{peint7}}
  \hspace{1em}
  \subcaptionbox{$I_6^{(4)}$\label{fig:I4l6}}{\showdiag{peint4}}
  \hspace{1em}
  \subcaptionbox{$I_7^{(4)}$\label{fig:I4l7}}{\showdiag{peint5}}
  \hspace{1em}
  \subcaptionbox{$I_8^{(4)}$\label{fig:I4l8}}{\showdiag{peint10}}
  \caption{Representative diagrams for integrals from sectors that we identified
    as potentially elliptic. Black solid lines represent massless propagators,
    green and red lines represent massive propagators with mass $m_b$ and $m_t$,
    respectively. For integrals $I_1^{(3)}$ and $I_{1,\dots,7}^{(4)}$ there
    exist corresponding integrals $\tilde{I}_1^{(3)}$ and
    $\tilde{I}_{1,\dots,7}^{(4)}$ that are related by swapping
    $m_b \leftrightarrow m_t$. Integral $I_8^{(4)}$ is symmetric under this
    transformation.}
  \label{fig:potentially-elliptic}
\end{figure}
Analysing the 197 sectors in this way lets us identify 15 sectors at four loops
that we have to analyse in more detail. Only eight of these sectors are
independent after taking into account permutations of $m_b$ and $m_t$, which do
not change the class of functions. This leaves us with eight sectors to analyse
in detail. Representative diagrams for each sector are shown in
\cref{fig:potentially-elliptic}. All of these sectors have Picard-Fuchs
operators of order two, three or four. In each case we find at most second-order
factors when trying to factorise the differential operators using
\texttt{DFactor}.

We further analyse the sectors by considering the maximal cut of their
integrals in $d=2$ dimensions since they fulfil homogeneous differential
equations \cite{Primo:2016ebd,Primo:2017ipr}. We work in loop-by-loop Baikov
representation \cite{Baikov:1996rk,Frellesvig:2017aai} as this allows us to
reuse building blocks from lower orders to construct the maximal cuts. In order
to identify non-polylogarithmic integrals, we investigate the geometric
structure of the maximal cuts. For example, integrals that depend on elliptic
curves show up as $\int \frac{\rmd z}{y}$ with $y^2 = P(z)$, where $P(z)$ is
a polynomial of degree 3 or 4. If $P(z)$ is a degree-$n$ polynomial, this would
correspond to a genus-$\left\lfloor \frac{n-1}{2} \right\rfloor$ curve.
Higher-dimensional structures, such as Calabi-Yau manifolds, would show up as
multivariate integrals over algebraic functions.

Before analysing these four-loop topologies, we collect some results about
maximal cuts of one- and two-loop integrals. We will use these results as
building blocks later in the analysis, and to help us interpret our findings.
At one loop, we are interested in bubble integrals, which we will denote
by $B_{ij}(q^2)$, where $i,j \in \{0,b,t\}$ denote the masses of the two
internal propagators and $q$ is the external momentum flowing through the
two-point integral. Up to irrelevant numerical prefactors, the maximal cut of
the one-loop bubble integrals are given by
\begin{align}
  \MaxCut B_{00}(q^2)
    &= \adjustbox{valign=m}{\tikz[very thick]{
         \draw pic[scale=0.8] (diagB00) at (0,0) {B00};
         \draw[cut] (0, 0.35) -- (0, 0.55);
         \draw[cut] (0,-0.35) -- (0,-0.55);
       }}
    \sim \frac{1}{q^2}
  \,, \\
  \MaxCut B_{b 0}(q^2)
    &= \adjustbox{valign=m}{\tikz[very thick]{
         \draw pic[scale=0.8] (diagBb0) at (0,0) {Bb0};
         \draw[cut] (0, 0.35) -- (0, 0.55);
         \draw[cut] (0,-0.35) -- (0,-0.55);
       }}
    \sim \frac{1}{q^2 - m_b^2}
  \,, \\
  \MaxCut B_{b b}(q^2)
    &= \adjustbox{valign=m}{\tikz[very thick]{
         \draw pic[scale=0.8] (diagBbt) at (0,0) {Bbb};
         \draw[cut] (0, 0.35) -- (0, 0.55);
         \draw[cut] (0,-0.35) -- (0,-0.55);
       }}
    \sim \frac{1}{\sqrt{q^2 [q^2 - 4 m_b^2]}}
  \,, \\
  \MaxCut B_{b t}(q^2)
    &= \adjustbox{valign=m}{\tikz[very thick]{
         \draw pic[scale=0.8] (diagBbt) at (0,0) {Bbt};
         \draw[cut] (0, 0.35) -- (0, 0.55);
         \draw[cut] (0,-0.35) -- (0,-0.55);
       }}
    \sim \frac{1}{\sqrt{[q^2 - (m_b - m_t)^2] [q^2 - (m_b + m_t)^2]}}
  \,.
\end{align}
We see that the maximal cut of the massless bubble $B_{00}(q^2)$ takes the form
of a massless propagator and that of the single-mass bubble $B_{b0}(q^2)$ looks
like a massive propagator. The maximal cuts of the equal-mass and unequal-mass
bubbles, $B_{bb}(q^2)$ and $B_{b t}(q^2)$, already contain square roots.
Considering the maximal cut of the equal-mass sunrise $S_{bbb}(q^2)$, we find
\cite{Remiddi:2016gno}\footnote{Here and in the following, we do not explicitly
specify the integration contour. Each independent contour yields a linearly
independent solution to the homogeneous differential equation.  However, the
specific contour is not relevant to our discussion.}
\begin{align}
  \MaxCut S_{bbb}(q^2)
    &= \adjustbox{valign=m}{\tikz[very thick]{
         \draw pic[scale=0.8] (diagSbbb) at (0,0) {Sbbb};
         \draw[cut] (0, 0.35) -- (0, 0.55);
         \draw[cut] (0,-0.10) -- (0, 0.10);
         \draw[cut] (0,-0.35) -- (0,-0.55);
       }}
  \notag \\
   &\sim \int \frac{\rmd z}{\sqrt{\left[z - \left(\sqrt{q^2} - m_b\right)^2\right]
         \left[z - \left(\sqrt{q^2} + m_b\right)^2\right]} \sqrt{z [z - 4 m_b^2]}}
  \label{eq:maxcut-Sbbb}
  \,.
\end{align}
In contrast to the previous examples, the maximal cut of the sunrise integral
is not just rational or algebraic functions but actually an elliptic integral.

Let us now consider $I_1^{(3)}$ (see \cref{fig:I3l1}), an integral which
contributes to the three-loop two-mass corrections to the $\rho$ parameter,
which was called $J_{8b}^{(3)}$ in Ref.~\cite{Grigo:2012ji}. We can think of
this integral as two one-loop bubble integrals that are inserted onto a
one-loop tadpole. This leads to the representation
\begin{align}
  I_1^{(3)} &\sim \int \rmd^d k \, B_{bt}(k^2) B_{bb}(k^2)
  \,.
\end{align}
The integration over $k$ corresponds to the integration over the loop momentum
of the one-loop tadpole. Inserting the building blocks for the maximal cuts of
the one-loop bubbles we discussed before and using the fact that the integrand
only depends on $k^2$ to perform the spherical part of the loop integration, we
find
\begin{align}
   \MaxCut I_1^{(3)}
     &\sim \int \frac{\rmd k^2}{\sqrt{[k^2 - (m_t - m_b)^2] \,
           [k^2 - (m_t+m_b)^2]} \, \sqrt{k^2 [k^2 - 4 m_b^2]}}
   \,.
\end{align}
Comparing this expression to \cref{eq:maxcut-Sbbb}, we see that this is exactly
the maximal cut of the on-shell ($q^2 = m_t^2$) two-loop sunrise integral.  We
will call the elliptic curve that appears here the \emph{sunrise curve} due to
this connection. Obviously, there exists a related integral $\tilde{I}_1^{(3)}$
that emerges upon swapping $m_b$ and $m_t$. The elliptic curve that shows up in
that integral will be called the \emph{sunrise twin curve}.

We are now ready to discuss a few examples at four-loop order. The first
integral, $I_3^{(4)}$ (see \cref{fig:I4l3}), can be read as a propagator, a
single-mass bubble and an equal-mass sunrise integral inserted on a one-loop
tadpole,
\begin{align}
  I_3^{(4)}
    &\sim \int \rmd^d k \, B_{b0}(k^2) \frac{1}{k^2 - m_t^2} S_{bbb}(k^2)
  \,.
\end{align}
Taking the maximal cut of this integral leads to
\begin{align}
  \MaxCut I_3^{(4)}
    &\sim \int \rmd k^2 \, \frac{\delta(k^2-m_t^2)}{k^2-m_b^2}
          \int \frac{\rmd z}{\sqrt{[z - (\sqrt{k^2}-m_b)^2] \,
          [z - (\sqrt{k^2} + m_b)^2]} \sqrt{z \, [z - 4 m_b^2]}}
  \,.
\end{align}
The $\delta$ distribution corresponds to the maximal cut of the single
propagator, the propagator-like factor $(k^2 - m_b^2)^{-1}$ is the maximal cut
of the single-mass bubble and the remaining integral over $z$ comes from
the maximal cut of the sunrise integral. The integral over the tadpole loop
momentum $k$ can be performed with the help of the $\delta$ distribution and we
arrive at
\begin{align}
  \MaxCut I_3^{(3)}
    &\sim \frac{1}{m_t^2 - m_b^2} \int \frac{\rmd z}{\sqrt{[z - (m_t-m_b)^2] \,
          [z - (m_t + m_b)^2]} \sqrt{z \, [z - 4 m_b^2]}}
  \,.
\end{align}
Inspecting the result reveals that this integral also depends on the sunrise
elliptic curve.

As a second example, we consider $I_6^{(4)}$ (see \cref{fig:I4l6}), an integral
that consists of three bubble integrals chained together in a loop. We call
integrals with this kind of topology \emph{necklace integrals} and we will
return to them below. The three bubbles are a massless bubble, an equal-mass
bubble and an unequal-mass bubble,
\begin{align}
  I_6^{(4)} &\sim \int \rmd^d k \, B_{00}(k^2) \, B_{bb}(k^2) \, B_{bt}(k^2)
  \,.
\end{align}
The maximal cut of this integral reads
\begin{align}
  \MaxCut I_6^{(4)}
    &\sim \int \frac{\rmd z}{z \sqrt{[z - (m_t-m_b)^2] \, [z - (m_t + m_b)^2]}
          \sqrt{z \, [z - 4 m_b^2]}}
  \,,
\end{align}
which, upon comparison with \cref{eq:maxcut-Sbbb}, again turns out to be
related to the sunrise curve.

Our next example looks very similar superficially, but turns out to exhibit
different behaviour. The integral $I_7^{(4)}$ (see \cref{fig:I4l7}) contains
two unequal-mass bubbles and one equal-mass bubble chained into a loop,
\begin{align}
  I_7^{(4)} &\sim \int \rmd^d k \, B_{bt}^2(k^2) \, B_{bb}(k^2)
  \,,
\end{align}
and its maximal cut takes the form
\begin{align}
  \MaxCut I_7^{(4)}
    &\sim \int \frac{\rmd z}{[z - (m_t-m_b)^2] \, [z - (m_t + m_b)^2]
          \sqrt{z \, [z - 4 m_b^2]}}
  \,.
\end{align}
Since the unequal-mass bubble appears twice, its contribution to the maximal
cut becomes squared, which turns the two branch points of the square root
appearing in the maximal cut of the bubble into simple poles of the full
integral. This reduces the degree of the polynomial under the square root from
four to two, which means that the maximal cut of this integral is not an
elliptic integral, but just an algebraic function.
Thus, integral $I_7^{(4)}$ serves as an example where the appearance of a
second-order factor in the Picard-Fuchs operator does not automatically imply
an elliptic solution.\footnote{We remark that the differential variable
influences whether we find an irreducible second-order factor the
Picard-Fuchs operator: trying to factorize the differential operator for
integral $I_7^{(4)}$ in $x^2 = m_b^2 / m_t^2$, as we first did, yields a
second-order factor, while the operator in $x$ actually fully factorises into
first-order factors. This is due to the restriction that the algorithm behind
the \texttt{DFactor} command works over the field of rational functions in the
differential variable. We chose to keep this example to point out this
important caveat.}

The final example we want to discuss here is $I_8^{(4)}$ (see \cref{fig:I4l8}),
an integral that consists of three bubbles, two equal-mass bubbles that however
each depend on different masses, and one unequal-mass bubble,
\begin{align}
  I_8^{(4)} &\sim \int \rmd^d k \, B_{tt}(k^2) \, B_{bb}(k^2) \, B_{bt}(k^2)
  \,,
\end{align}
whose maximal cut is
\begin{align}
  \MaxCut I_8^{(4)}
    &\sim \int \frac{\rmd z}{z \, \sqrt{[z - 4 m_t^2]} \sqrt{[z - 4 m_b^2]}
          \sqrt{[z - (m_t-m_b)^2] \, [z - (m_t + m_b)^2]}}
  \,.
\end{align}
While this integral also features a degree-$4$ polynomial under the square root
and yields an elliptic integral, the elliptic curve that appears is related
neither to the sunrise curve nor to the sunrise twin curve.

Overall we find from our analysis that three different elliptic curves appear
in the integrals: the sunrise curve, the sunrise twin curve, and the new curve
that shows up in integral $I_8^{(4)}$. The curves are defined by $y^2 = P_i(z)$
with
\begin{align}
  P_1(z) &= z \, [z-4 m_b^2] \, [z - (m_t-m_b)^2] \, [z - (m_t+m_b)^2]
    \,, && \text{(sunrise curve)} \\
  P_2(z) &= z \, [z - (m_t-m_b)^2] \, [z - (m_t+m_b)^2] \, [z-4 m_t^2]
    \,, && \text{(sunrise twin curve)} \\
  P_3(z) &= [z-4 m_b^2] \, [z - (m_t-m_b)^2] \, [z - (m_t+m_b)^2] \, [z-4 m_t^2]
    \,. && \text{(new curve)}
\end{align}
We find the following classification
\begin{align*}
  I_1^{(3)},~I_{1,\dots,6}^{(4)} &\rightsquigarrow P_1(z)
  \,, &
  \tilde{I}_1^{(3)},~\tilde{I}_{1,\dots,6}^{(4)} &\rightsquigarrow P_2(z)
  \,, \\
  I_7^{(4)},~\tilde{I}_7^{(4)} &\rightsquigarrow \text{non-elliptic}
  \,, &
  I_8^{(4)} &\rightsquigarrow P_3(z)
  \,.
\end{align*}
We have checked by computing the $j$~invariant of the curves that they are not
isomorphic to each other, see, e.g., Ref.~\cite{Weinzierl:2022eaz} for details.
Moreover, we verified that the curves are not isogenic by computing the periods
numerically for randomly chosen values of the masses.

The maximal cut analysis in loop-by-loop Baikov representation has the
additional advantage that it reveals an intuitive physical interpretation of
the branch points of the elliptic curves. The branch points of the square roots
in the maximal cuts of the one-loop bubbles correspond to the thresholds and
pseudothresholds of these integrals. Since the final integration of the
four-loop integrals corresponds to the square of the loop momentum $k$ that
flows through the bubble integrals, we can interpret the branch points of the
elliptic curves also as thresholds or pseudothresholds in the internal loop
momentum space.  Given two different masses, there are in total five different
(pseudo)thresholds and, indeed, each of the three curves uses four of those
five thresholds as its branch points.

Thus far, we have only analysed the maximal cuts of the integrals that
contribute to the four-loop $\rho$ parameter. In order to gain a fuller picture
of the class of functions that are necessary to express the result, we also
have to take into account how the inhomogeneous parts of the differential
equations couple different integrals to each other. In the three-loop
calculation, two elliptic curves appeared, but sectors featuring different
curves were never coupled to each other and individual integrals always only
depended on a single elliptic curve. Therefore, it is possible to express the
result in terms of elliptic multiple polylogarithms
\cite{Abreu:2019fgk}.\footnote{Alternatively, it is possible to express the
result in terms of iterated integrals over kernels involving hypergeometric
${}_2 F_1$ functions \cite{Ablinger:2017bjx,Blumlein:2018aeq}.} However, in
the calculation at hand, we find that integrals which depend on the three
different elliptic curves we identified are coupled to each other by the
differential equations. This implies that elliptic multiple polylogarithms may
not be not sufficient to express the results.

\begin{figure}
  \centering
  \subcaptionbox{Four-loop genus-2 necklace\label{fig:genus2}}[5cm]{%
    \begin{tikzpicture}[very thick]
      \draw pic[scale=1.0] {genus2};
    \end{tikzpicture}
  }
  \subcaptionbox{$L$-loop genus-$(L-1)$ necklace\label{fig:genusL-1}}[5cm]{%
    \begin{tikzpicture}[very thick,scale=0.6]
      \begin{feynman}
        \vertex (v1) at (0:2);
        \vertex (v2) at (60:2);
        \vertex (v3) at (120:2);
        \vertex (v4) at (180:2);
        \vertex (v5) at (240:2);
        \vertex (v6) at (300:2);
        \diagram*{
          (v1) --[plain,mb,bend left] (v2),
          (v1) --[plain,mb,bend right] (v2),
          (v2) --[plain,mt,bend left] (v3),
          (v2) --[plain,mb,bend right] (v3),
          (v3) --[plain,mt,bend left] (v4),
          (v3) --[plain,m3,bend right] (v4),
          (v4) --[plain,m3,bend left] (v5),
          (v4) --[plain,mb,bend right] (v5)
        };
      \end{feynman}
      \draw[dotted] (-10:2) -- (-19:2);
      \draw[dotted] (250:2) -- (259:2);
    \end{tikzpicture}
  }
  \caption{Example diagrams that give rise to Feynman integrals that depend on
    higher-genus curves.}
  \label{fig:higher-genus-diags}
\end{figure}
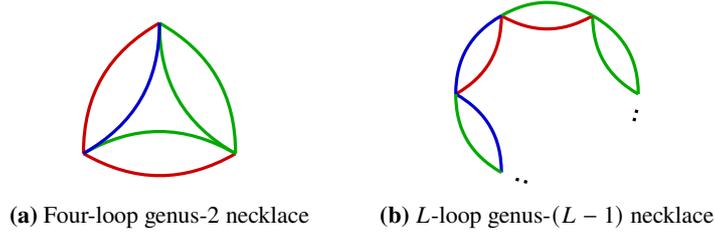
Finally, let us comment on how generalized versions of some of the diagrams
that contribute to the $\rho$ parameter give rise to integrals over
\emph{hyperelliptic} curves. In particular, the diagrams in
\crefrange{fig:I4l6}{fig:I4l8} can be thought of as degenerations of the
four-loop necklace diagram shown in \cref{fig:genus2}. When we have particles
with only two different masses appearing in this integral, the number of branch
points is constrained by combinatorics. That is, the pseudothresholds of
equal-mass bubbles necessarily agree ($m_b - m_b = m_t - m_t = 0$). Therefore,
we can only construct curves with at most four non-repeated roots. However, if
we allow for three different masses, we could combine two different
unequal-mass bubbles and an equal-mass bubble, see \cref{fig:genus2}, to
generate a degree-$6$ polynomial,
\begin{align}
  y^2 &= z \, [z - 4 m_b^2] \, [z - (m_t-m_b)^2] \, [z - (m_t + m_b)^2]
         [z - (m_t-m_3)^2] \, [z - (m_t + m_3)^2]
  \,,
\end{align}
and thus a genus-2 curve.

Going beyond four-loop order, we can even construct $L$-loop integrals that
give rise to integrals over a genus-$(L-1)$ curve as their maximal cuts. Similar
to the construction above, we chain $(L-1)$ one-loop bubbles into a loop, while
ensuring that the mass assignment is such that none of the thresholds or
pseudothresholds coincide, see \cref{fig:genusL-1}. The maximal cut of an
integral constructed in this way will depend on a genus-$(L-1)$ curve. We note
that the minimal number of masses that are necessary to ensure distinct branch
points grows rather slowly. These integrals represent an interesting class of
examples and can serve as toy models for studying integrals that depend on
higher-genus curves and show up as soft limits of integrals that contribute to
the Standard Model.

\section{Conclusions and outlook}
We have reported on our ongoing calculation of the two-mass QCD contributions to
the four-loop $\rho$ parameter. This is an interesting testing ground for which
special functions appear in Standard Model observables. We find three distinct
elliptic curves that contribute to the integrals and observe that the
differential equations couple integrals that depend on different elliptic
curves. This poses challenges for finding analytic closed form solutions to
these Feynman integrals.

Starting from the expressions for the weak boson self energies in terms of
master integrals, it is straightforward to calculate a numerical value for these
four-loop corrections. However, since the tadpole integrals are universal
building blocks that can be useful also for other calculations, we will
calculate series representations that yield results for arbitrary values of the
masses. In the long term, it would be interesting to investigate how to
effectively and efficiently express the tadpole integrals in terms of special
functions with known properties. Finally, we also plan to study necklace
integrals with more than two masses as examples for Feynman integrals that
depend on higher-genus curves.

\section*{Acknowledgements}
The work of AM has been supported by the Royal Society grant
URF\textbackslash{}R1\textbackslash{}221233.
The work of BP has received funding from the European's Union Horizon 2020
research and innovation programme LoopAnsatz (grant agreement number 896690).

\bibliographystyle{JHEPM}
\bibliography{lit}

\end{document}